\documentclass[aps,prr,amsmath,amssymb,longbibliography,lengthcheck,showpacs,superscriptaddress,floatfix]{revtex4-1}
\usepackage{graphicx}
\usepackage{color}

\begin{document}

\title{Anharmonic properties of vibrational excitations in amorphous solids}

\author{Hideyuki Mizuno}
\email{hideyuki.mizuno@phys.c.u-tokyo.ac.jp} 
\affiliation{Graduate School of Arts and Sciences, The University of Tokyo, Tokyo 153-8902, Japan}
\author{Masanari Shimada}
\affiliation{Graduate School of Arts and Sciences, The University of Tokyo, Tokyo 153-8902, Japan}
\author{Atsushi Ikeda}
\affiliation{Graduate School of Arts and Sciences, The University of Tokyo, Tokyo 153-8902, Japan}

\date{\today}

\begin{abstract}
Understanding the vibrational and thermal properties of amorphous solids is one of the most discussed and long-standing issues in condensed matter physics.
Recent works have made significant steps towards understanding harmonic vibrational states.
In particular, it has been established that quasi-localized vibrational modes emerge in addition to phonon-like vibrational modes.
In this work, we study the anharmonic properties of these vibrational modes.
We find that vibrational modes exhibit anharmonicities that induce particle rearrangements and cause transitions to different states.
These anharmonicities are distinct from those in crystals, where particle rearrangements never occur.
Remarkably, for both the phonon modes and quasi-localized modes, the vibrational modes exhibit strong anharmonicities, and the induced particle rearrangements are always localized in space and are composed of $1$ to $1000$ particles.
Our findings contribute to the understanding of low-temperature thermal properties, for which anharmonic vibrations are crucial.
\end{abstract}

\maketitle

\section{Introduction}
Amorphous solids exhibit vibrational and thermal properties that are markedly different from those of crystals~\cite{Phillips_1981,Elliott_1990}.
Recent numerical simulations have led to progress in the understanding of the harmonic vibrational states in amorphous solids.
In crystals, harmonic vibrational states are well established as phonons~\cite{Ashcroft_1976,Kittel_1996}.
On the other hand, in amorphous solids, quasi-localized vibrational~(QLV) modes emerge in addition to phonon-like vibrational modes~\cite{Lerner_2016,Mizuno_2017,Shimada_2018,Lerner_2018,Wang_2019,Ikeda_2019}.
In the QLV mode, particles in some regions vibrate greatly, while other particles vibrate much less.
The vibrational density of states~(vDOS) of phonon modes, $g_\text{phonon}(\omega)$~($\omega$ is frequency), follows the Debye law, $g_\text{phonon}(\omega) = A_D \omega^2$~($A_D$ is the Debye level), as does that of phonons in crystals, whereas the vDOS of QLV modes, $g_\text{QLV}(\omega)$, follows the non-Debye scaling law $g_\text{QLV}(\omega) = \alpha \omega^4$~($\alpha$ is constant)
\footnote{
The exponent $4$ of $g_\text{QLV}(\omega)$ may be changed to around $3$ to $4$ through the preparation protocol of the system~\cite{Lerner_2018}.
}.
The existence of QLV modes enhances scattering in phonon transport~\cite{Mizuno_2018,Wang2_2019,Moriel_2019}, which in particular induces Rayleigh scattering even in the zero-temperature limit, as observed by simulations~\cite{Monaco2_2009,Marruzzo_2013,Mizuno_2014} and experiments~\cite{Masciovecchio_2006,Monaco_2009,Baldi_2010}.

The total vDOS of vibrational modes is therefore described as $g(\omega) = g_\text{phonon}(\omega) + g_\text{QLV}(\omega)$
\footnote{
In the thermodynamic limit of $N \to \infty$, phonon modes and QLV modes can hybridize~\cite{Bouchbinder_2018}.
Although effects of the hybridization are not yet understood, we expect that the total vDOS is still described by $g(\omega) = g_\text{phonon}(\omega) + g_\text{QLV}(\omega)$.
}.
If we suppose that there are harmonic vibrations in the system, then we predict the specific heat at low temperatures~\cite{Ashcroft_1976,Kittel_1996} to be $C(T) = C_\text{phonon}(T) + C_\text{QLV}(T) = C_D T^3 + \beta T^5$~($T$ is temperature, $C_D$ is the Debye level of specific heat, and $\beta$ is a constant related to $\alpha$ in $g_\text{QLV}(\omega)$).
However, this $T$ dependence of $C(T)$ cannot explain the experimental observation that the specific heat linearly depends on $T$~\cite{Phillips_1981,Elliott_1990,Zeller_1971}.
This result certainly demonstrates that anharmonic vibrations are crucial even at low $T$.
This situation is totally different from the case of crystals.
In crystals, anharmonicities become small or even negligible at low $T$, and as a result, the specific heat can be well described by harmonic vibrations without anharmonicities~\cite{Ashcroft_1976,Kittel_1996}.

For amorphous solids, another type of anharmonicity has been proposed, the so-called two-level system~(TLS)~\cite{Anderson_1972,Phillips_1972,Phillips_1987,Galperin_1989}.
In a TLS, the system transits between two different states with an energy difference, which may be realized through quantum tunneling effects at low $T$.
Importantly, TLSs induce particle rearrangements, which is never the case for anharmonicities in crystals.
If we suppose the existence of many different TLSs with different energy scales, then we can indeed explain the linear $T$ dependence of the specific heat~\cite{Anderson_1972,Phillips_1972,Phillips_1987,Galperin_1989}.
Subsequently, TLS theory has been extended to the soft potential model~\cite{Karpov_1983,Karpov_1985,Buchenau_1991,Buchenau_1992,Gurevich_2003,Gurevich_2005,Gurevich_2007}, which attempts to describe vibrational anomalies, including the QLV, TLS, and the excess low-frequency vibrational modes~(so-called boson peak)~\cite{buchenau_1984,Yamamuro_1996,Mizuno2_2013,Kabeya_2016}, in an unified framework.
Experiments~(e.g., Refs.~\cite{Graebner_1986,Queen_2013,Perez-Castaneda_2014,Perez-Castaneda2_2014}) as well as numerical simulations~(e.g., Refs.~\cite{Weber_1985,Heuer_1993,Heuer_1996,Reinisch_2004,Middleton_2001,Damart_2018}) have attempted to detect the TLSs and clarify their nature, e.g., the statistics of parameters characterizing the TLSs~(such as the distance and energy difference between two states), transition paths connecting two states, or the density of TLSs.

On another front, a recent work~\cite{Shimada2_2018} demonstrated that the QLV modes exhibit unstable vibrations in a localized region.
Additionally, Refs.~\cite{Gartner_2016,Gartner2_2016,Wijtmans_2017} disentangled the localized defects from extended vibrational modes, which can be used to predict the location of plastic instabilities.
These results might motivate us to expect that the transition between two states in TLSs could be induced by unstable vibrations in the QLV modes or localized defects embedded in the extended vibrational modes.
However, the relationship between TLSs and QLV modes or localized defects remains to be solved.
In addition to these works, Ref.~\cite{Xu_2010} conducted a pioneering study on the anharmonic properties of vibrational modes and suggested that the low-frequency modes can exhibit strong anharmonicities.

Considering the current status described above, the present work studies the anharmonic properties of vibrational excitations based on the recently advanced understanding of vibrational modes.
Since systems at finite $T$ are excited along the vibrational modes by thermal energy, it could be of primary importance to understand the anharmonic properties of vibrational modes.
We focus on each vibrational mode and forcibly excite it to measure its anharmonic properties.
We attempt to address the following questions.
(i) Can the vibrational modes exhibit anharmonicities that induce particle rearrangements and cause transitions to different states?
(ii) If so, what are properties of the particle rearrangements and the transitions?
(iii) What are the differences in the anharmonicities between phonon modes and QLV modes?
We also discuss TLSs in relation to what we find regarding the anharmonicities of vibrational modes.

\section{Method}
\subsection{System description}~\label{system}
In the present work, we perform molecular dynamics~(MD) simulations.
Our numerical system is composed of monodispersed $N$ point particles of mass $m$ in a three-dimensional cubic box of length $L$ and volume $V \equiv L^3$.
We implement periodic boundary conditions in all directions.
We employ two types of pairwise, interparticle potentials, a harmonic potential and the Lennard-Jones~(LJ) potential, as described below.

\subsubsection{Harmonic potential system}~\label{harmonic}
Particles interact via the following harmonic, pairwise potential:
\begin{equation}~\label{harmonic}
\phi_\text{HA}(r) = \frac{\epsilon}{2} \left( 1 - \frac{r}{\sigma} \right)^2 H(\sigma -r),
\end{equation}
where $r$ is the distance between two particles and $H(x)$ is the Heaviside step function: $H(x) = 1$ for $x \ge 0$ and $H(x) = 0$ for $x<0$.
$\sigma$ is the particle diameter, and $\epsilon$ is the energy scale.
We measure physical quantities in units of mass $m$, length $\sigma$, and energy $\epsilon$.
The frequency, $\omega$, and temperature, $T$, are measured via $\sqrt{ \epsilon / (m \sigma^2) }$ and $\epsilon/k_B$~($k_B$ is the Boltzmann constant), respectively.
This harmonic potential was originally proposed for modeling granular materials, emulsions, foams, etc.~\cite{van_Hecke_2009}.
However, here, we employ this system as the simplest model of glass~\cite{Wyart_2005,Wyart_2010,DeGiuli_2014}.

Throughout this work, the packing fraction is fixed as $\varphi \equiv (\pi \sigma^3/6)(N/V) \approx 0.73$~\cite{Mizuno_2017}.
We start with a random configuration at infinite temperature, $T = \infty$, and instantaneously quench the system to zero temperature, $T=0$, where the pressure is $p = 5 \times 10^{-2}$.
Here, we use the FIRE algorithm~\cite{Bitzek_2006} for quenching the system~(minimizing the potential energy).
To access the low-frequency vibrational modes, several different system sizes, from $N = 16000$ to $512000$, are simulated.

\subsubsection{Lennard-Jones potential system}~\label{lj}
Particles interact via the LJ pairwise potential:
\begin{equation}~\label{lj}
\phi_\text{LJ}(r) = 4\epsilon \left[ \left( \frac{\sigma}{r}\right)^{12} - \left(\frac{\sigma}{r} \right)^{6} \right],
\end{equation}
where $r$ is the distance between two particles, $\sigma$ is the particle diameter, and $\epsilon$ is the energy scale.
We truncate the potential at a cut-off distance $r = r_c \equiv 2.5 \sigma$.
The potential and the force~(the first derivative of the potential) are then shifted to zero at $r = r_c$, as in Ref.~\cite{Shimada_2018}.
As in the case of the harmonic potential, we use units of mass $m$, length $\sigma$, and energy $\epsilon$.

Throughout this work, the number density is fixed as $\hat{\rho} \equiv N/V \approx 0.997$~\cite{Shimada_2018}.
We first equilibrate the system in the normal liquid state at $T=2$ and then quench the system to zero temperature, $T=0$, where the pressure is around zero, $p = 0$.
We employ several different system sizes, from $N=4000$ to $128000$.

\subsection{Vibrational mode}~\label{method:ha}
We now obtain the $T=0$ configuration of particles, i.e. the inherent structure, for two types of amorphous systems.
In our previous works on the harmonic potential system~\cite{Mizuno_2017} and the LJ potential system~\cite{Shimada_2018}, we studied the harmonic vibrational states in a wide range of frequency.
Specifically, we diagonalized the Hessian matrix~(second derivative of the potential) and obtained the $3N-3$ vibrational modes, where we discarded three zero-frequency, translational modes.
Each vibrational mode $k$~($k=1,2,...,3N-3$) is characterized by the eigenfrequency, $\omega^k$, and the eigenvector, $\mathbf{e}^k_i$, of each particle $i$~($i=1,2,...,N$).
Here, the eigenvector is orthonormalized as $\sum_{i=1}^N \mathbf{e}^{k}_i \cdot \mathbf{e}^{l}_i = \delta_{kl}$~($\delta_{kl}$ is the Kronecker delta).

As explained in the Introduction, there exist two types of vibrational modes in the low-frequency regime, phonon modes and QLV modes~\cite{Mizuno_2017,Shimada_2018}.
The vDOS, which is defined as $g(\omega) = \left[ 1/(3N-3) \right] \sum_{k=1}^{3N-3} \delta (\omega-\omega^k)$~($\delta(x)$ is the Dirac delta function), is described as the sum of those of phonon modes and QLV modes: $g(\omega) = g_\text{phonon}(\omega) + g_\text{QLV}(\omega) = A_D \omega^2 + \alpha \omega^4$.
We note that such vibrational properties~(phonon modes and QLV modes) appear in the low-$\omega$ regime, below the boson peak frequency $\omega_\text{BP}$~\cite{Mizuno_2017,Shimada_2018}.
$\omega_\text{BP}$ is defined as the frequency at which the reduced vDOS, $g(\omega)/\omega^2$, takes a maximum~\cite{buchenau_1984,Yamamuro_1996,Mizuno2_2013,Kabeya_2016}.
To distinguish these two types of modes, we have measured two order parameters: the phonon order parameter ${O}^k$ and participation ratio ${P}^k$.
Below, we briefly explain how to calculate these two order parameters.
For more details, please see our previous works~\cite{Mizuno_2017,Shimada_2018}.

\subsubsection{Phonon order parameter ${O}^k$}
The phonon order parameter, $O^k$, evaluates the extent to which eigenvector $\mathbf{e}^k_i$~($i=1,2,...,N$) of mode $k$ is similar to phonon vibrations.
We first define the displacement vectors of phonon modes as $\mathbf{u}^{\mathbf{q},\alpha}_i = \mathbf{s}_\alpha(\hat{\mathbf{q}}) \exp\left( {\text{i}\mathbf{q}\cdot\mathbf{r}_{0i}} \right)/{\sqrt{N}}$, where $\mathbf{r}_{0i}$ is the position of particle $i$ in the inherent structure, $\mathbf{q}$ is the wave vector, and $\hat{\mathbf{q}} = \mathbf{q}/\left| \mathbf{q} \right|$.
$\alpha$ denotes one longitudinal ($\alpha=L$) and two transverse ($\alpha=T_1,T_2$) phonon modes.
$\mathbf{s}_\alpha(\hat{\mathbf{q}})$ is a unit vector that represents the direction of polarization: $\mathbf{s}_{L}(\hat{\mathbf{q}}) = \hat{\mathbf{q}}$ (longitudinal) and $\mathbf{s}_{T_1}(\hat{\mathbf{q}}) \cdot \hat{\mathbf{q}}= \mathbf{s}_{T_2}(\hat{\mathbf{q}}) \cdot \hat{\mathbf{q}} = 0$ (transverse).

We then define the phonon order parameter, $O^k$, as
\begin{equation} \label{phononorder}
\begin{aligned}
O^k & = \sum_{\mathbf{q},\alpha;\ O^k_{\mathbf{q},\alpha} \ge N_m/(3N-3)} O^k_{\mathbf{q},\alpha}, \\
O^k_{\mathbf{q},\alpha} & = \left| \sum_{i=1}^N \mathbf{u}^{\mathbf{q},\alpha}_i \cdot \mathbf{e}^k_i \right|^2,
\end{aligned}
\end{equation}
where $N_m = 100$ is employed; however, we confirm that our results and conclusions do not depend on the choice of the value of $N_m$.
As extreme cases, $O^k = 1$ for an ideal phonon mode and $O^k = 0$ for a mode considerably different from phonon modes.

\subsubsection{Participation ratio ${P}^k$}
The participation ratio, $P^k$, quantitatively measures the extent of localization, which has often been employed in early works~\cite{Schober_1991,mazzacurati_1996,Taraskin_1999}.
Given the eigenvector $\mathbf{e}^{k}_i$~($i=1,2,...,N$) of mode $k$, its participation ratio, ${P}^k$, is calculated as
\begin{equation}~\label{eq:participation}
{P}^k \equiv \frac{1}{N} \left[ \sum_{i=1}^{N} \left( \mathbf{e}^k_i \cdot \mathbf{e}^k_i \right)^2 \right]^{-1}.
\end{equation}
${P}^k$ quantifies the fraction of particles that participate in the vibrations, and thus, $N P^k$ indicates the number of participating particles.
As extreme cases, ${P}^k = 1$~($N P^k = N$) for an ideal mode in which all the constituent particles vibrate equally, ${P}^k = 1/N \ll 1$~($N P^k = 1$) for an ideal mode involving only one particle, and ${P}^k = 2/3$ for an ideal plane wave.

\begin{figure}[t]
\centering
\includegraphics[width=0.49\textwidth]{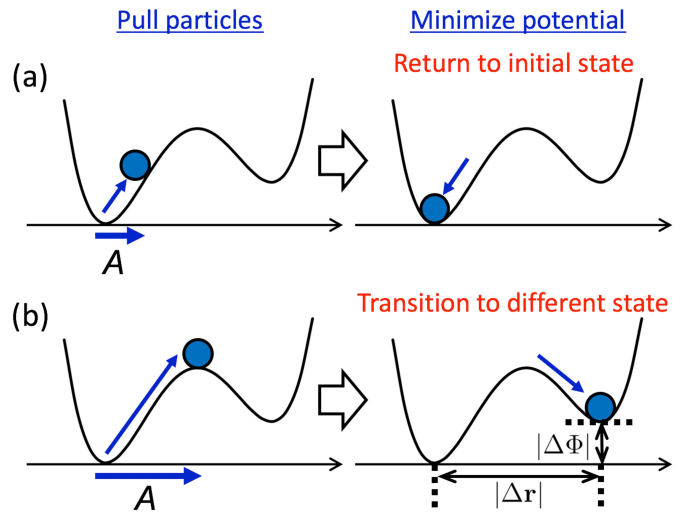}
\vspace*{0mm}
\caption{\label{fig1}
{Measurement of the anharmonicity of the vibrational mode.}
In the figure, we illustrate the potential energy landscape, $\Phi(\{\mathbf{r}_i\})$.
We first pull the particles along the direction of mode $k$~($\mathbf{e}^k_i$) by $A$ and next minimize the potential energy.
Then, we will obtain one of two consequences: (a)~the system returns to the initial state~(initial inherent structure), or (b)~the system transitions to a different state~(different inherent structure).
In (b), the distance $|\Delta \mathbf{r}|$ and the potential energy difference $|\Delta \Phi|$ between the initial state and the state after transition characterize the properties of the transition.
}
\end{figure}

\begin{figure}[t]
\centering
\includegraphics[width=0.49\textwidth]{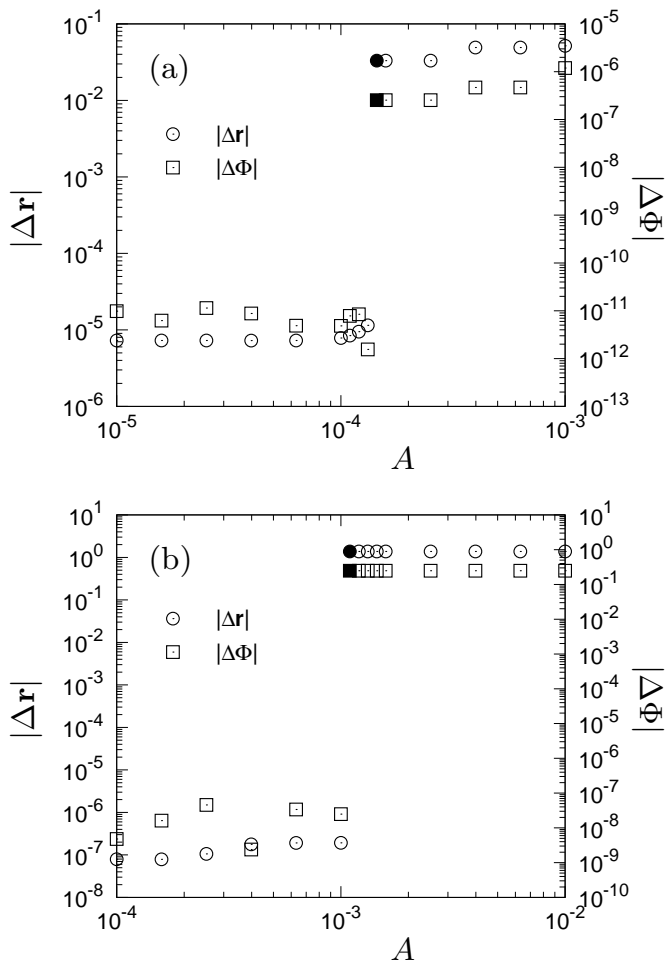}
\vspace*{0mm}
\caption{\label{fig2}
{Example of measurement of the anharmonicity of the vibrational mode.}
Plot of the distance, $|\Delta \mathbf{r}|$, and the potential energy difference, $|\Delta \Phi|$, as functions of the extent of particle pulling, $A$.
(a) Harmonic potential system and (b) LJ potential system.
As we increase $A$, we see clear jumps in $|\Delta \mathbf{r}|$ and $\Delta \Phi$ at some value of $A = A_c$~(as indicated by closed symbols), at which the system transitions to a different state.
}
\end{figure}

\subsection{Anharmonicity of the vibrational mode}~\label{method:anha}
In the present work, we study the anharmonic properties of the vibrational modes.
To measure the anharmonicities, we follow a procedure performed in previous work~\cite{Xu_2010}.
First, we focus on vibrational mode $k$ and forcibly excite it: we start with the particles in the inherent structure and pull them along the direction of mode $k$ by changing the position of particle $i$~($i=1,2,...,N$) as
\begin{equation}~\label{eq:pull}
\mathbf{r}_i = \mathbf{r}_{0i} + A \sqrt{N} \mathbf{e}^k_i,
\end{equation}
where $\mathbf{r}_{0i}$ and $\mathbf{r}_i$ are the initial position of particle $i$ before pulling~(i.e., position in the inherent structure) and the present position of particle $i$ after pulling, respectively~(see Fig.~\ref{fig1}).
$A$ measures the extent of particle pulling, as $A =\sqrt{(1/N)\sum_{i=1}^N (\mathbf{r}_i-\mathbf{r}_{0i})^2}$, i.e., the average value of the displacements of particles.
We note that the factor $\sqrt{N}$ in Eq.~(\ref{eq:pull}) is necessary for $A$ to be independent of the system size $N$~\cite{Schreck_2011}~
\footnote{
If the target mode $k$ is \textit{truly} localized in space where only some portion of particles vibrate and the other particles exhibit exactly zero vibrations, then the factor $\sqrt{N}$ should be erased for $A$ to be independent of the system size.
}.
After pulling the particles, we next minimize the potential energy of the system, $\Phi(\{\mathbf{r}_i\}) \equiv \sum_{i<j} \phi(r_{ij})$.
Then, we will obtain one of two consequences, as shown in Fig.~\ref{fig1}: (a)~the system returns to the initial state~(initial inherent structure), or (b)~the system transitions to a different state~(different inherent structure).

To see whether the system (a)~returns or (b)~transitions, we monitor the distance between the initial state and the state after potential minimization, defined as
\begin{equation}~\label{eq:dis}
\left| \Delta \mathbf{r} \right| = \sqrt{\sum_{i=1}^N (\mathbf{r}_{1i} - \mathbf{r}_{0i})^2 },
\end{equation}
where $\mathbf{r}_{1i}$ is the position of particle $i$ in the state after potential minimization.
We also monitor the potential energy difference
\begin{equation}~\label{eq:enedis}
\left| \Delta \Phi \right| = \left| \Phi(\{ \mathbf{r}_{1i} \}) - \Phi(\{ \mathbf{r}_{0i} \}) \right|.
\end{equation}
Figure~\ref{fig2} shows the distance, $|\Delta \mathbf{r}|$, and the potential energy difference, $|\Delta \Phi|$, as functions of $A$, for both the harmonic potential and LJ potential systems.
When $A$ is small, $|\Delta \mathbf{r}|$ and $|\Delta \Phi|$ both stay at zero, which means that (a) the system returns to the initial state.
As we increase the value of $A$, we see discontinuous jumps in $|\Delta \mathbf{r}|$ and $|\Delta \Phi|$ at some values of $A = A_c$~(as indicated by closed symbols in Fig.~\ref{fig2}), at which (b) the system transitions to a different state.
Note that $|\Delta \mathbf{r}|$ and $|\Delta \Phi|$ exhibit jumps at exactly the same value of $A = A_c$.
This result means that when the vibrational mode $k$ is excited by $A_c$, anharmonicity can emerge via induction of particle rearrangement, causing the transition to a different state.
$A_c$ therefore measures the extent of anharmonicity of mode $k$: a smaller $A_c$ indicates a stronger anharmonicity.

Here, we make a note on the LJ potential system.
For several vibrational modes, we can pull the particles by a large amount $A$ along the mode without transition.
In this case, some pairs of particles largely overlap, and the potential tends to diverge.
We disregarded these cases.

We also emphasize that anharmonicities in the crystalline solids do not induce particle rearrangement causing changes in the inherent structure~(periodic lattice structure)~\cite{Ashcroft_1976,Kittel_1996}.
On the other hand, amorphous solids exhibit anharmonicities that do induce particle rearrangements and changes in the inherent structure.
These anharmonicities can emerge due to the amorphous structure.

\section{Results}
We analyzed the anharmonic properties of vibrational modes by using different system sizes, from $N = 16000$ to $512000$ for the harmonic potential system and from $N=4000$ to $128000$ for the LJ potential system.
Below, we show data for these different system sizes all together, which verifies the absence of system size effects.

\begin{figure}[t]
\centering
\includegraphics[width=0.49\textwidth]{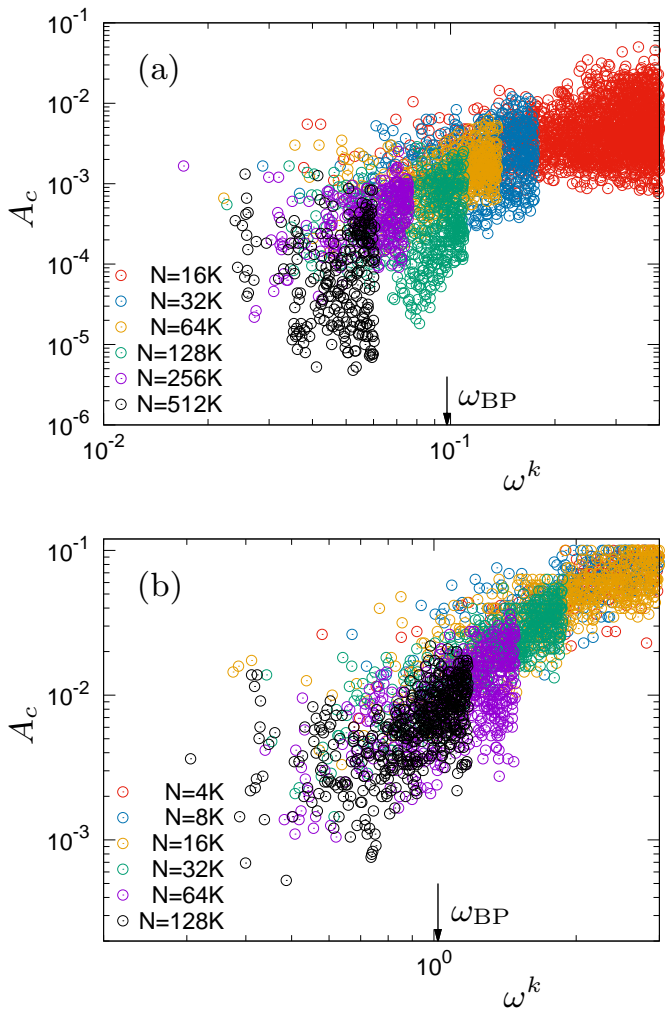}
\vspace*{0mm}
\caption{\label{fig3}
{Extent of anharmonicity of vibrational modes.}
Plot of the extent of anharmonicity, $A_c$, as a function of the eigenfrequency of the excited vibrational mode, $\omega^k$.
(a) Harmonic potential system and (b) LJ potential system.
We plot data from different system sizes using different color symbols.
Additionally, we indicate the boson peak frequency, $\omega_\text{BP}$, by an arrow.
}
\end{figure}

\begin{figure}[t]
\centering
\includegraphics[width=0.49\textwidth]{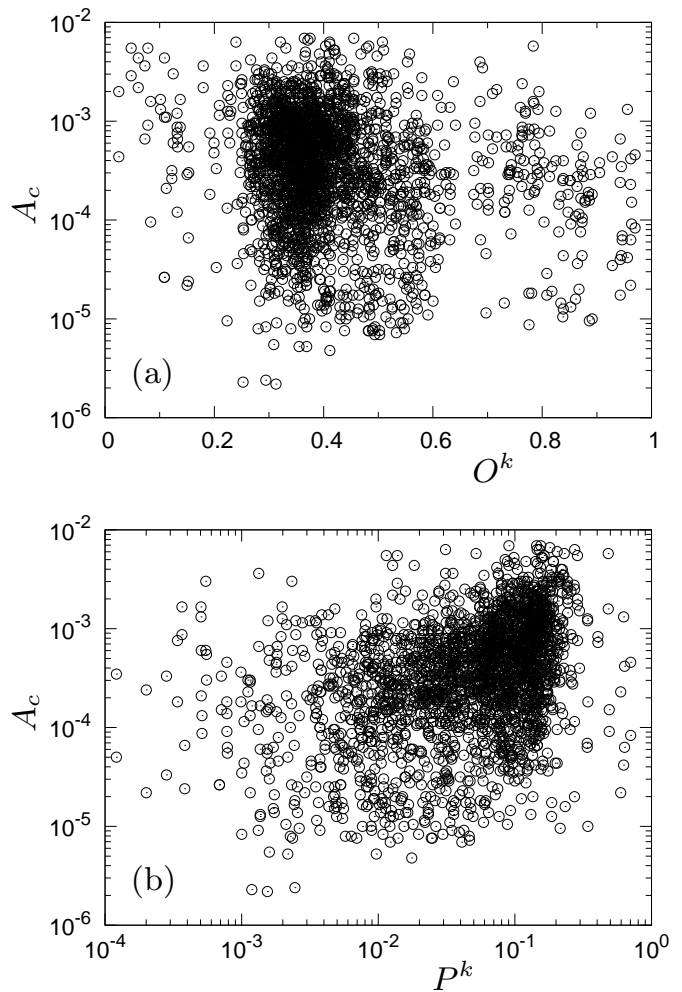}
\vspace*{0mm}
\caption{\label{fig4}
{Correlation between anharmonicities and nature of excited vibrational modes in the harmonic potential system.}
Plot of the extent of anharmonicity, $A_c$, as a function of the phonon order parameter, $O^k$, in (a) and the participation ratio, $P^k$, in (b).
We plot data for vibrational modes below the boson peak frequency, $\omega_\text{BP}$.
}
\end{figure}

\subsection{Extent of anharmonicity of vibrational modes}~\label{result:exanha}
Figure~\ref{fig3} plots the extent of anharmonicity, $A_c$, as a function of the eigenfrequency of the excited vibrational mode, $\omega^k$, for the harmonic potential system in (a) and LJ potential system in (b).
In the figure, we indicate the boson peak frequency, $\omega_\text{BP}$, by an arrow.
We see that $A_c$ becomes small as $\omega^k$ decreases, i.e., the lower frequency mode shows a stronger anharmonicity.
This has already been observed for the harmonic potential system in previous work~\cite{Xu_2010}; here, we confirmed the same result for the LJ potential system.
Remarkably, $A_c$ takes quite small values, particularly below $\omega_\text{BP}$: $A_c \sim 10^{-6}$ to $10^{-2}$~(of the particle size) for the harmonic potential system and $A_c \sim 10^{-4}$ to $10^{-1}$ for the LJ potential system.
Possibly, $A_c$ goes to zero in the zero-frequency limit of $\omega^k \to 0$.

In addition, Figure~\ref{fig4} plots $A_c$ as a function of the phonon order parameter, $O^k$, in (a) and the participation ratio, $P^k$, in (b), for the modes below $\omega_\text{BP}$ in the harmonic potential system~(we also obtain similar results for the LJ potential system).
We observe no apparent correlations between the values of $A_c$ and $O^k$ or $P^k$, i.e., between the extent of anharmonicity and nature of the excited vibrational modes.
Particularly, there are no differences in the extent of anharmonicity between the phonon modes with large $O^k$ and $P^k$ and the QLV modes with small $O^k$ and $P^k$.
We therefore conclude that irrespective of phonon modes or QLV modes, the low-frequency vibrational modes exhibit similar extents of anharmonicity that induce particle rearrangement and cause transitions to different states.

In Appendix~\ref{expansion}, we analyze the anharmonicities of the eigenmodes by expanding the potential energy, $\Phi(\{\mathbf{r}_i\})$, around the inherent structure $\{ \mathbf{r}_{0i} \}$ in terms of $A \sqrt{N}$ up to the third-order term.
We obtain an analytical expression for $A_c$ in Eq.~(\ref{acexpression}) as the saddle point in the expanded potential energy.
The value of $A_c$ in Eq.~(\ref{acexpression}) can be estimated using the eigenvalues ${\omega^k}^2$~(the second-order derivative of the potential) and the third-order derivative of the potential at $\{\mathbf{r}_i\} = \{ \mathbf{r}_{0i} \}$.
For both the harmonic potential and LJ potential systems, the value of $A_c$ measured by Eq.~(\ref{acexpression}) is orders of magnitude larger than that in Fig.~\ref{fig3} at fixed $\omega^k$~(see Fig.~\ref{fig13}).
We can also see that the $A_c$ value from Eq.~(\ref{acexpression}) is strongly correlated with the values of $O^k$ and $P^k$, which is in contrast to the observation in Fig.~\ref{fig4}~(see Fig.~\ref{fig14}).
These results clearly demonstrate that simply expanding the potential energy landscape around the inherent structure cannot correctly estimate the anharmonicities, which is due to the rather complex shape of the potential energy landscape~\cite{Weber_1985,Heuer_1993,Heuer_1996,Reinisch_2004,Munro_1999,Middleton_2001,Damart_2018}.

\begin{table}[t]
\centering
\renewcommand{\arraystretch}{1.1}
\begin{tabular}{c|c||c|c|c}
\hline
\hline
\ Harmonic \ &\ $N$ \  & $32000$ & $128000$ & $512000$ \\
\cline{2-5}
             &\ $T_c$ \ &\ $6.2\times 10^{-7}$ \ &\ $2.8\times 10^{-7}$ \ &\ $3.8\times 10^{-9}$ \\
\hline
\hline
\ LJ         &\ $N$ \   & $8000$ & $32000$ & $128000$ \\
\cline{2-5}
             &\ $T_c$ \ &\ $1.1\times 10^{-1}$ \ &\ $1.9\times 10^{-2}$ \ &\ $6.8\times 10^{-3}$ \\
\hline
\hline
\end{tabular}
\caption{
Onset temperature of the anharmonicities, $T_c$.
The value of $T_c$ is reported for both the harmonic potential and LJ potential systems and for different system sizes $N$.
} \label{table1}
\end{table}

From the data in Fig.~\ref{fig3}, we can discuss the onset temperature of the present anharmonicities when the system is equilibrated in the classical manner without any quantum effect.
When the system is equilibrated at temperature $T$, the thermal energy of $T/2$ is distributed to each vibrational mode according to the equipartition law of energy.
If no anharmonicities are induced, then $T < \left[ \omega^k A_c(\omega^k) \right]^2 N$ should be satisfied for all modes $k$~(where we explicitly denote $A_c$ as a function of $\omega^k$).
Thus, when $T$ exceeds $T_c \equiv \left\{ \left[ \omega^k A_c(\omega^k) \right]_\text{min} \right\}^2 N$, the anharmonicities are induced by the mode that takes the minimum value of $\left[ \omega^k A_c(\omega^k) \right]$, denoted as $\left[ \omega^k A_c(\omega^k) \right]_\text{min}$, and particle rearrangements occur.
Table~\ref{table1} reports the values of $T_c$ for different system sizes $N$.
$T_c$ decreases with increasing $N$, i.e., anharmonicities are more easily induced in larger systems.
For large systems, $T_c$ takes quite small values of $T_c \sim 10^{-9}$ for the harmonic potential system~($N=512000$) and $T_c \sim 10^{-3}$ for the LJ potential system~($N=128000$).
Therefore, tiny thermal fluctuations can cause the anharmonicities to induce particle rearrangements.
We note that for the harmonic potential system, the value of $T_c$ is much less than the onset temperature where the effects of anharmonicities become visible in the macroscopic quantities~(moment of vDOS) in Ref.~\cite{Ikeda_2013}.
We also note that our value of onset temperature $T_c$ possibly goes to zero in the thermodynamic limit, $N \to \infty$.
Although the method to measure anharmonicities is different, this result is consistent with the argument in Ref.~\cite{Schreck_2011} that amorphous systems can be inherently anharmonic.

\begin{figure}[t]
\centering
\includegraphics[width=0.49\textwidth]{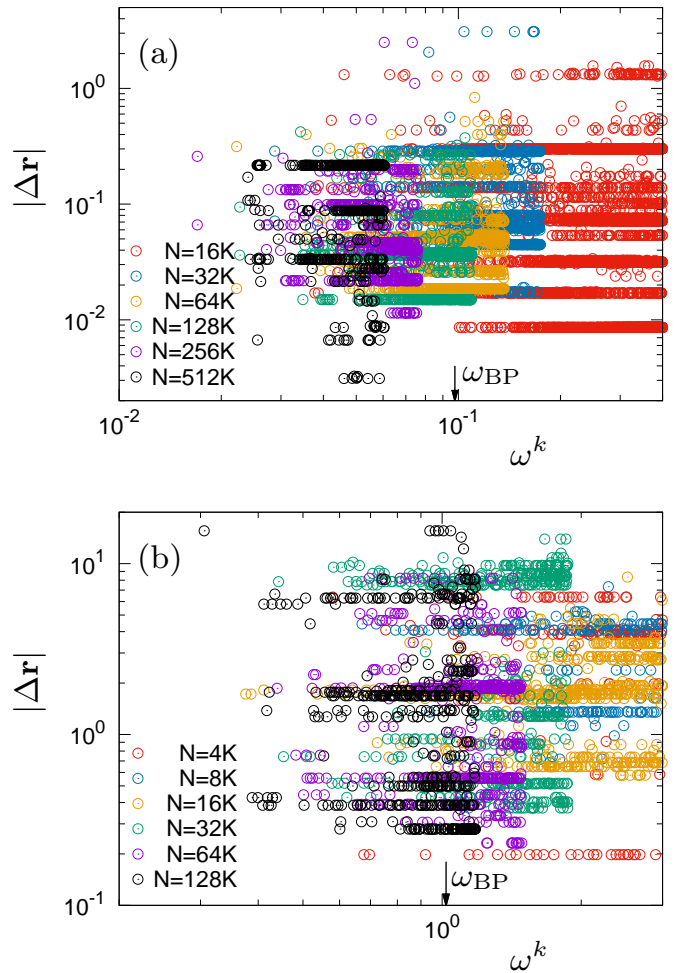}
\vspace*{0mm}
\caption{\label{fig5}
{Distance between the initial state and the state after transition.}
Plot of $|\Delta \mathbf{r}|$ in Eq.~(\ref{eq:dis}) as a function of the eigenfrequency of the excited vibrational mode, $\omega^k$.
(a) Harmonic potential system and (b) LJ potential system.
We plot data from different system sizes using different color symbols.
}
\end{figure}

\begin{figure}[t]
\centering
\includegraphics[width=0.49\textwidth]{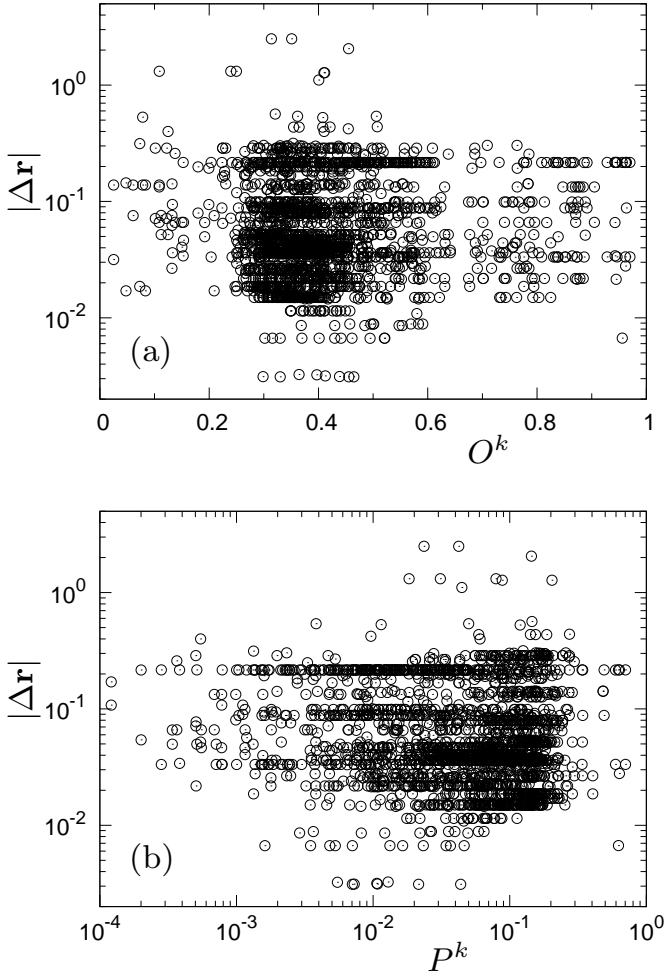}
\vspace*{0mm}
\caption{\label{fig6}
{Correlation between transitions and nature of excited vibrational modes in the harmonic potential system.}
Plot of $|\Delta \mathbf{r}|$ as a function of the phonon order parameter, $O^k$, in (a) and the participation ratio, $P^k$, in (b).
We plot data for vibrational modes below the boson peak frequency, $\omega_\text{BP}$.
}
\end{figure}

\begin{figure}[t]
\centering
\includegraphics[width=0.49\textwidth]{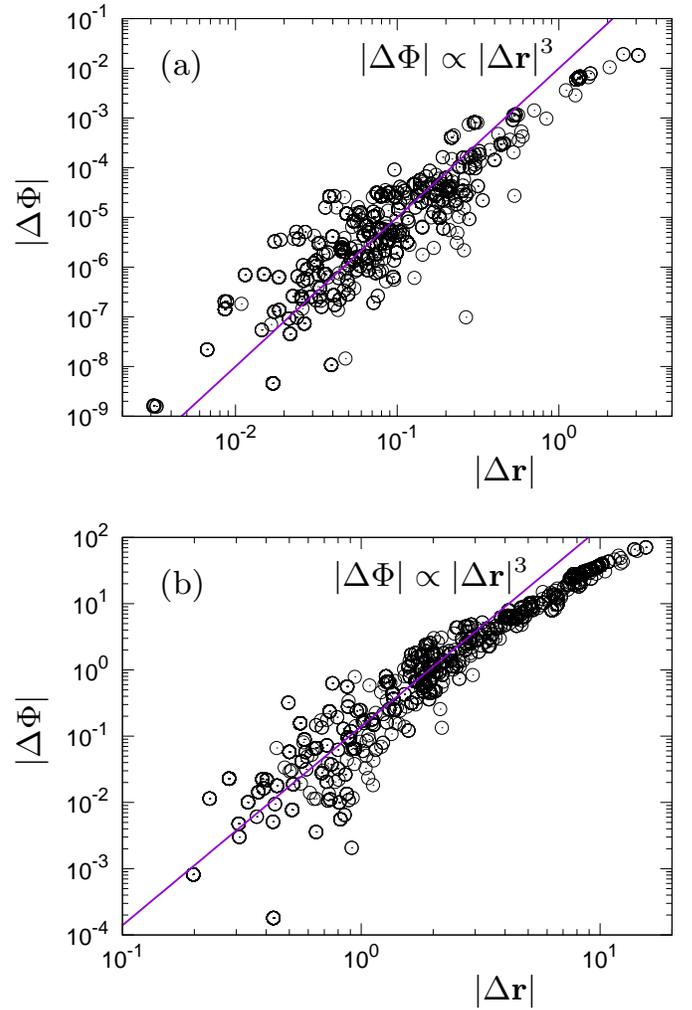}
\vspace*{0mm}
\caption{\label{fig7}
{Distance versus potential energy difference between the initial state and the state after transition.}
Plot of $|\Delta \mathbf{r}|$ versus $|\Delta \Phi|$ for all of the studied transitions.
(a) Harmonic potential system and (b) LJ potential system.
The solid line indicates $|\Delta \Phi| \propto |\Delta \mathbf{r}|^3$.
}
\end{figure}

\subsection{Distance and potential energy difference between the initial state and state after transition}~\label{result:tranha}
We next study distributions of the distance, $|\Delta \mathbf{r}|$ in Eq.~(\ref{eq:dis}), and the potential energy difference, $|\Delta \Phi|$ in Eq.~(\ref{eq:enedis}), between the initial state and the state after transition.
We illustrate the values of $|\Delta \mathbf{r}|$ and $|\Delta \Phi|$ in Fig.~\ref{fig1}(b).
Figure~\ref{fig5} plots $|\Delta \mathbf{r}|$ as a function of $\omega^k$ for the harmonic potential system in (a) and the LJ potential system in (b).
We see that $|\Delta \mathbf{r}|$ does not apparently depend on $\omega^k$, which takes values of $|\Delta \mathbf{r}| \sim 10^{-3}$ to $10^0$ for the harmonic potential system and $|\Delta \mathbf{r}| \sim 10^{-1}$ to $10^1$ for the LJ potential system.
Importantly, we clearly observe that many different modes exhibit the same $|\Delta \mathbf{r}|$, i.e., many different modes cause the same transition.
Therefore, there are some particular transitions that are frequently caused by many different vibrational excitations, and the number of states after the transition is much fewer than that of vibrational modes~(i.e., $3N-3$).

Here, we make a note on finite system size effects.
From Fig.~\ref{fig5}, we confirm no apparent size effects on the values of $|\Delta \mathbf{r}|$~(we also confirm no size effects on $|\Delta \Phi|$).
As we will discuss in the next section~\ref{result:ipanha}, the induced particle rearrangements are always localized in space.
This localized nature in transitions leads to the fact that $|\Delta \mathbf{r}|$ and $|\Delta \Phi|$ are independent of the system size~$N$.

We also study correlations between the transitions and nature of excited vibrational modes.
Figure~\ref{fig6} plots $|\Delta \mathbf{r}|$ as a function of $O^k$ in (a) and $P^k$ in (b) for the modes below $\omega_\text{BP}$ in the harmonic potential system.
The figure does not show any correlation of $|\Delta \mathbf{r}|$ with $O^k$ or $P^k$: the induced transitions are not related to the nature of the excited vibrational modes.
In particular, the induced transitions are not correlated with whether the phonon modes or the QLV modes are excited. 
This result implies that the frequently induced transitions~(particle rearrangements) have roots in the shape of the energy landscape~\cite{Weber_1985,Heuer_1993,Heuer_1996,Reinisch_2004,Munro_1999,Middleton_2001,Damart_2018} or the structural properties of the amorphous inherent structure~\cite{Hua_2018,Hua_2019,Tanaka_2019}.

Furthermore, we plot $|\Delta \mathbf{r}|$ versus $|\Delta \Phi|$ for all the studied transitions in Fig.~\ref{fig7}.
We see a clear relation between $|\Delta \mathbf{r}|$ and  $|\Delta \Phi|$, which is roughly estimated as $|\Delta \Phi| \propto |\Delta \mathbf{r}|^3$: the large rearrangements, $|\Delta \mathbf{r}|$, induce a large energy difference, $|\Delta \Phi|$.
$|\Delta \Phi|$ takes values of $|\Delta \Phi| \sim 10^{-9}$ to $10^{-2}$ for the harmonic potential system and $|\Delta \Phi| \sim 10^{-4}$ to $10^2$ for the LJ potential system.

\begin{figure}[t]
\centering
\includegraphics[width=0.49\textwidth]{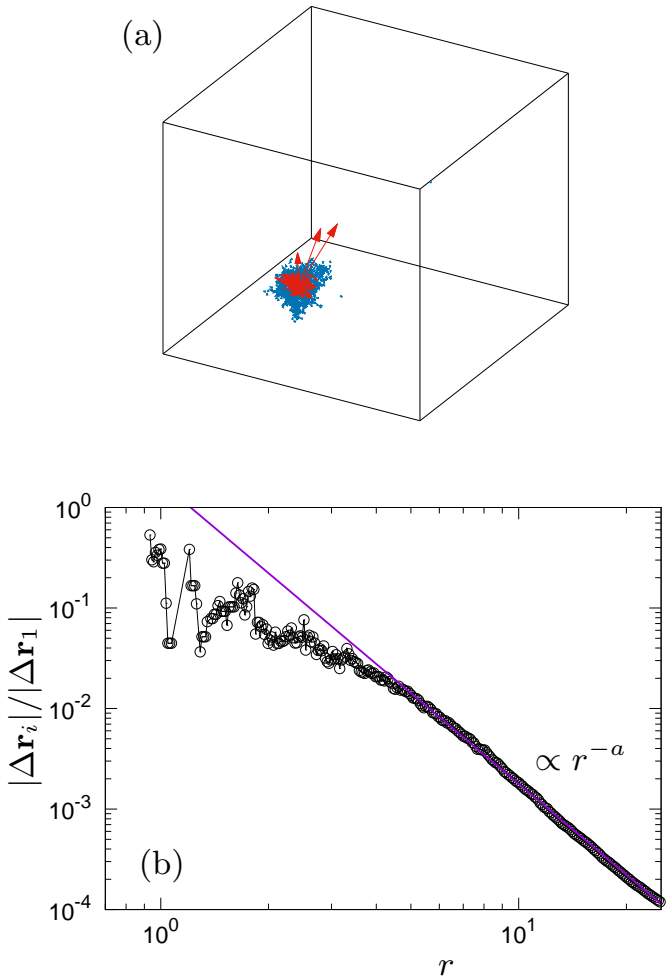}
\vspace*{0mm}
\caption{\label{fig8}
{Induced particle rearrangements in the harmonic potential system.}
The system size is $N=512000$.
(a) Visualization of the particle rearrangement field, $\Delta \mathbf{r}_i \equiv \mathbf{r}_{1i} - \mathbf{r}_{0i}$, as a function of $\mathbf{r}_{0i}$ in three-dimensional space for a representative transition.
We plot $\Delta \mathbf{r}_i \times 200$ for the largest $2000$ particles, where red arrows indicate the largest $100$ values.
(b) Plot of the normalized displacement of particle $i$, $|\Delta \mathbf{r}_i|/|\Delta \mathbf{r}_1|$, as a function of the distance $r$ from particle $1$ with the largest displacement, $|\Delta \mathbf{r}_1|$.
The tail at large distances $r$ behaves as a power law of $|\Delta \mathbf{r}_i| \propto r^{-a}$, with $a=3$ for the present case.
We confirm that the exponent, $a$, can take values of approximately $2.5$ to $3$.
}
\end{figure}

\begin{figure}[t]
\centering
\includegraphics[width=0.49\textwidth]{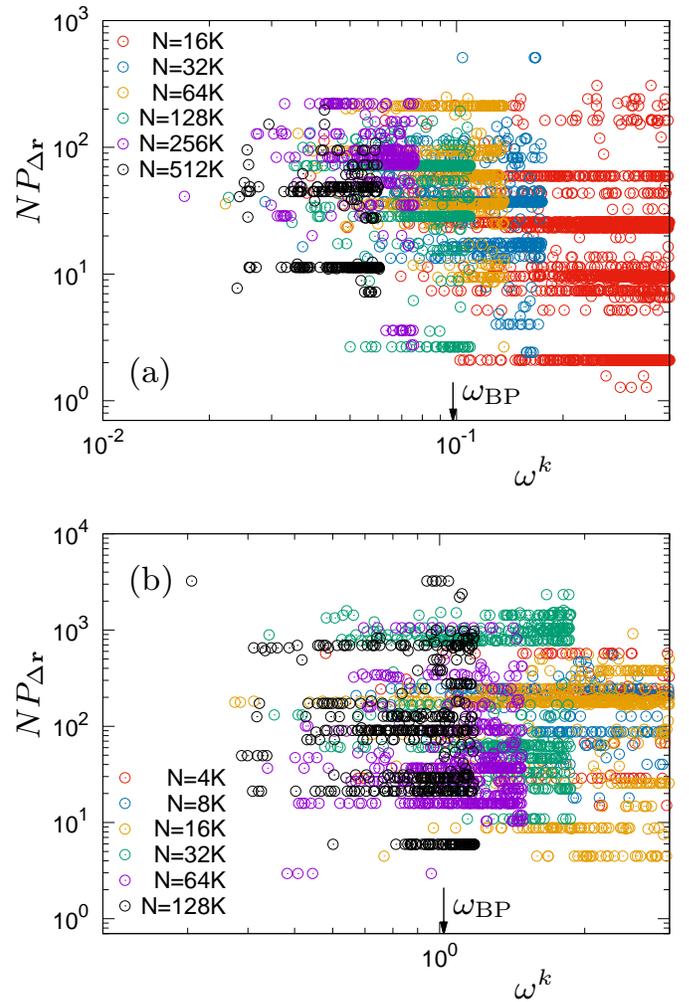}
\vspace*{0mm}
\caption{\label{fig9}
{Number of particles that participate in the particle rearrangement.}
Plot of $N {P_{\Delta \mathbf{r}}}$ in Eq.~(\ref{eq:number}) as a function of the eigenfrequency of the excited vibrational mode, $\omega^k$.
(a) Harmonic potential system and (b) LJ potential system.
We plot data from different system sizes using different color symbols.
Additionally, we indicate the boson peak frequency, $\omega_\text{BP}$, by the arrow.
}
\end{figure}

\begin{figure}[t]
\centering
\includegraphics[width=0.49\textwidth]{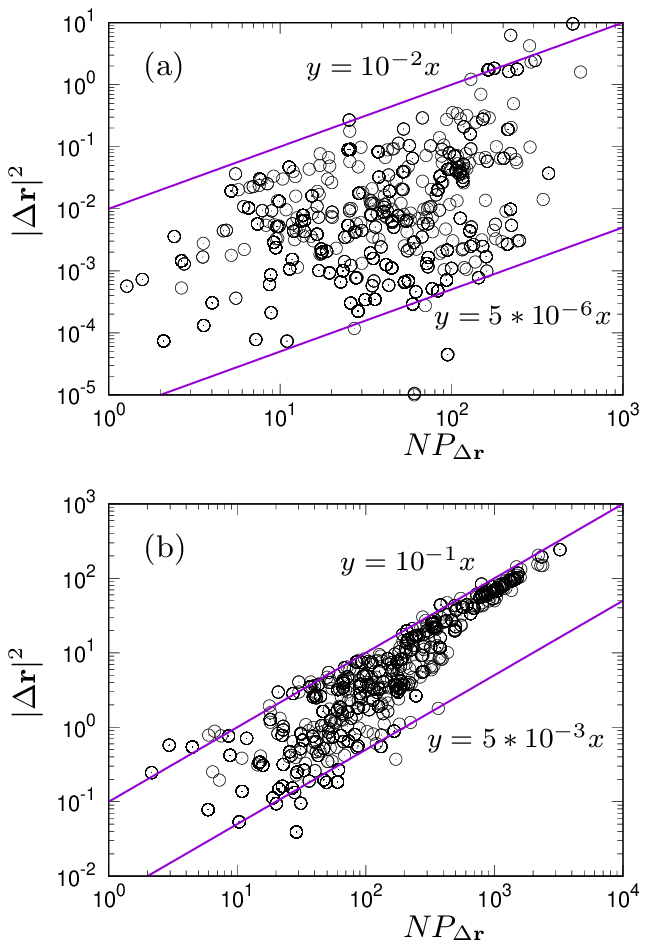}
\vspace*{0mm}
\caption{\label{fig10}
{Number of particles that participate in the particle rearrangement versus displacement of particles.}
Plot of number of particles, $N {P_{\Delta \mathbf{r}}}$, versus the square displacement, $\Delta \mathbf{r}^2$, for all the studied transitions~(particle rearrangements).
(a) Harmonic potential system and (b) LJ potential system.
The two solid lines roughly indicate the lower and upper bounds described by the lines of $\Delta \mathbf{r}^2 \propto N {P_{\Delta \mathbf{r}}}$.
}
\end{figure}

\subsection{Profile of induced particle rearrangement}~\label{result:ipanha}
In the previous sections, we have revealed that the low-frequency mode exhibits quite strong anharmonicities that induce particle rearrangement and cause transitions to different states.
We next study the profile of the induced particle rearrangement.
Let us denote the vector field of particle rearrangement as $\Delta \mathbf{r}_i \equiv \mathbf{r}_{1i} - \mathbf{r}_{0i}$, where we recall that $\mathbf{r}_{0i}$ and $\mathbf{r}_{1i}$ are the initial position of particle $i$ and the position after the transition, respectively.
Figure~\ref{fig8}(a) visualizes the vector field, $\Delta \mathbf{r}_i$, as a function of $\mathbf{r}_{0i}$ in three-dimensional space for a representative transition in a harmonic potential system with $N=512000$.
We immediately recognize that the rearrangement is highly localized in space.
In the figure, we observe a single localized region, but we also observe multiple localization regions in some cases.
We confirm the localization of particle rearrangement for all the studied transitions and for both the harmonic potential and LJ potential systems.

Figure~\ref{fig8}(b) plots the normalized displacement of particle $i$, $|\Delta \mathbf{r}_i|/|\Delta \mathbf{r}_1|$, as a function of the distance $r$ from particle $1$ with the largest displacement, $|\Delta \mathbf{r}_1|$.
We see that the tail at large distances $r$ behaves as a power law of $|\Delta \mathbf{r}_i| \propto r^{-a}$, with $a=3$ for the case shown in the figure.
We confirm that $a$ takes values of approximately $2.5$ to $3$.
This result indicates that the localized region is surrounded by some particle displacement field characterized by $|\Delta \mathbf{r}_i| \propto r^{-a}$, with $a \approx 2.5$ to $3$.
We note that the elastic deformation produces the field of $|\Delta \mathbf{r}_i| \propto r^{-2}$, i.e., $a=2$, whereas the present far-fields exhibit steeper decay~(with $a \approx 2.5$ to $3$).
We speculate that this steeper decay may be attributed to the nonlinear nature~\cite{Coulais_2014} or plastic nature~~\cite{Maloney_2006,Manning_2011} of rearrangement of particles, which could be a future subject.
We also note that a similar profile of particle displacements is observed in the elastic response to local forcing~\cite{Leonforte_2005,Lerner_2014,Karimi_2015} and the QLV modes~\cite{Lerner_2016}, which, however, show an elastic-deformation field with exponent $a=2$.

Furthermore, we measure the number of particles that participate in the rearrangement, which can be measured in the same manner as the $N P^k$ of the vibrational mode in Eq.~(\ref{eq:participation}):
\begin{equation}~\label{eq:number}
N P_{\Delta \mathbf{r}} \equiv \left[ \sum_{i=1}^{N} \left( \Delta \mathbf{r}_i \cdot \Delta \mathbf{r}_i \right) \right]^2 \left[ \sum_{i=1}^{N} \left( \Delta \mathbf{r}_i \cdot \Delta \mathbf{r}_i \right)^2 \right]^{-1}.
\end{equation}
Figure~\ref{fig9} plots $N P_{\Delta \mathbf{r}}$ as a function of $\omega^k$.
We observe that the value of $N P_{\Delta \mathbf{r}}$ does not apparently depend on the system size, $N$, i.e., $N P_{\Delta \mathbf{r}} \propto N^0$ or $P_{\Delta \mathbf{r}} \propto N^{-1}$.
This observation means that the particle rearrangements are localized in space.
From the figure, we can estimate that the number of participating particles is $1$ to $1000$ for the harmonic potential system and roughly the same or slightly more for the LJ potential system.

Additionally, Figure~\ref{fig10} plots $N {P_{\Delta \mathbf{r}}}$ versus the square displacement, $\Delta \mathbf{r}^2$~(see Eq.~(\ref{eq:dis}) for the formulation of $\Delta \mathbf{r}$).
We observe that data points exist roughly between the two lines of $\Delta \mathbf{r}^2 = 5*10^{-6} (N {P_{\Delta \mathbf{r}}})$ and $\Delta \mathbf{r}^2 = 10^{-2} (N {P_{\Delta \mathbf{r}}})$ for the harmonic potential system and $\Delta \mathbf{r}^2 = 5*10^{-3} (N {P_{\Delta \mathbf{r}}})$ and $\Delta \mathbf{r}^2 = 10^{-1} (N {P_{\Delta \mathbf{r}}})$ for the LJ potential system.
From this result, we can estimate the value of $\sqrt{\Delta \mathbf{r}^2/(N {P_{\Delta \mathbf{r}}})}$, which measures the displacement of each particle averaged over the participating particles, as $\sqrt{\Delta \mathbf{r}^2/(N {P_{\Delta \mathbf{r}}})} = 2*10^{-3}$ to $10^{-1}$~(of the particle size) for the harmonic potential system and $\sqrt{\Delta \mathbf{r}^2/(N {P_{\Delta \mathbf{r}}})} = 7*10^{-2}$ to $3*10^{-1}$ for the LJ potential system.
To summarize the results in this subsection, the induced particle rearrangement is spatially localized, where $1$ to $1000$ particles are displaced by roughly $10^{-3}$ to $10^{-1}$ of the particle size for each.

\begin{figure}[t]
\centering
\includegraphics[width=0.49\textwidth]{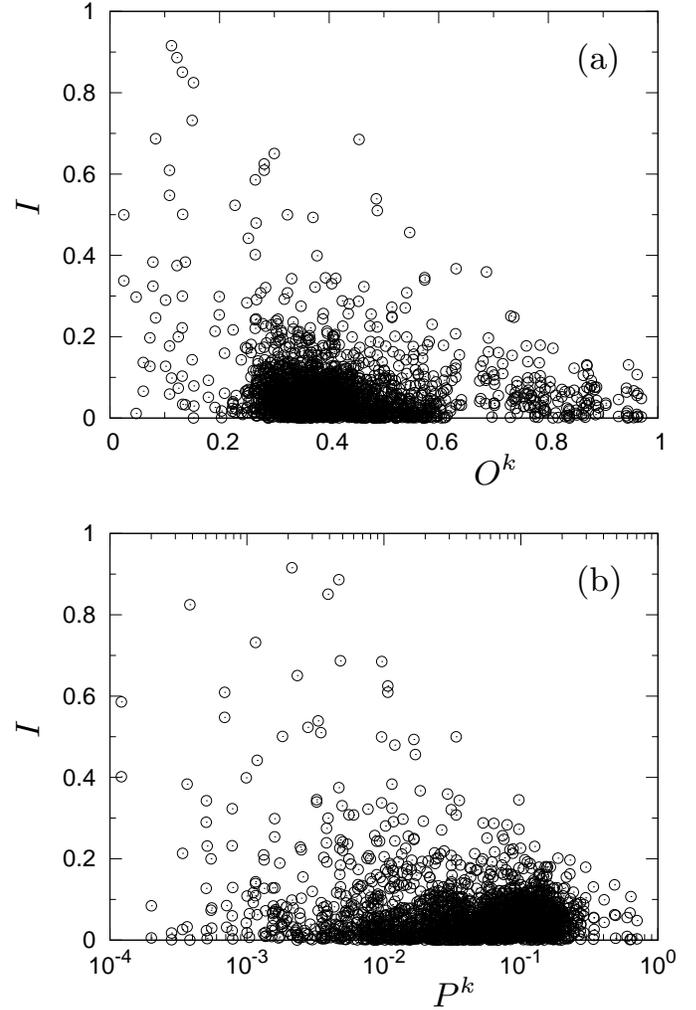}
\vspace*{0mm}
\caption{\label{fig11}
{Correlation in directions between particle rearrangements and vibrational modes in the harmonic potential system.}
We plot the value of the normalized inner product~$I$ between $\Delta \mathbf{r}_i$ and $\mathbf{e}_i^k$, defined in Eq.~(\ref{eq:inner}), as a function of the phonon order parameter, $O^k$, in (a) and the participation ratio, $P^k$, in (b).
Data are plotted for the low-frequency modes below the boson peak frequency, $\omega_\text{BP}$.
}
\end{figure}

\subsection{Correlation in directions between particle rearrangements and vibrational modes}~\label{result:coranha}
We next study the correlation in directions between particle rearrangements and excited vibrational modes, which can be measured by the normalized inner product between the displacement field, $\Delta \mathbf{r}_i$, and the eigenvector of excited mode $k$, $\mathbf{e}_i^k$:
\begin{equation}~\label{eq:inner}
I \equiv \left| \sum_{i=1}^N \Delta \mathbf{r}_i \cdot \mathbf{e}_i^k \right| \left[ \sum_{i=1}^N \Delta \mathbf{r}_i^2 \right]^{-1/2}.
\end{equation}
Note that $I$ takes values from $0$ to $1$.
$I = 1$ means that $\Delta \mathbf{r}_i$ and $\mathbf{e}_i^k$ are in the same direction: the particle rearrangements occur perfectly along the direction of mode $k$.
In contrast, $I = 0$ means that $\Delta \mathbf{r}_i$ and $\mathbf{e}_i^k$ are orthogonal to each other: the particle rearrangements occur in a totally different direction from that of mode $k$.

Figure~\ref{fig11} plots $I$ versus $O^k$ in (a) and $P^k$ in (b).
We see some correlations~(relatively large values of $I$) for some QLV modes with small values of $O^k$ and $P^k$.
However, the correlations are generally weak; even the QLV modes exhibit rather weak correlations with the particle rearrangements.
We therefore conclude that although some particle rearrangements occur along the QLV modes, they do not occur along the vibrational modes in general.

\begin{figure}[t]
\centering
\includegraphics[width=0.49\textwidth]{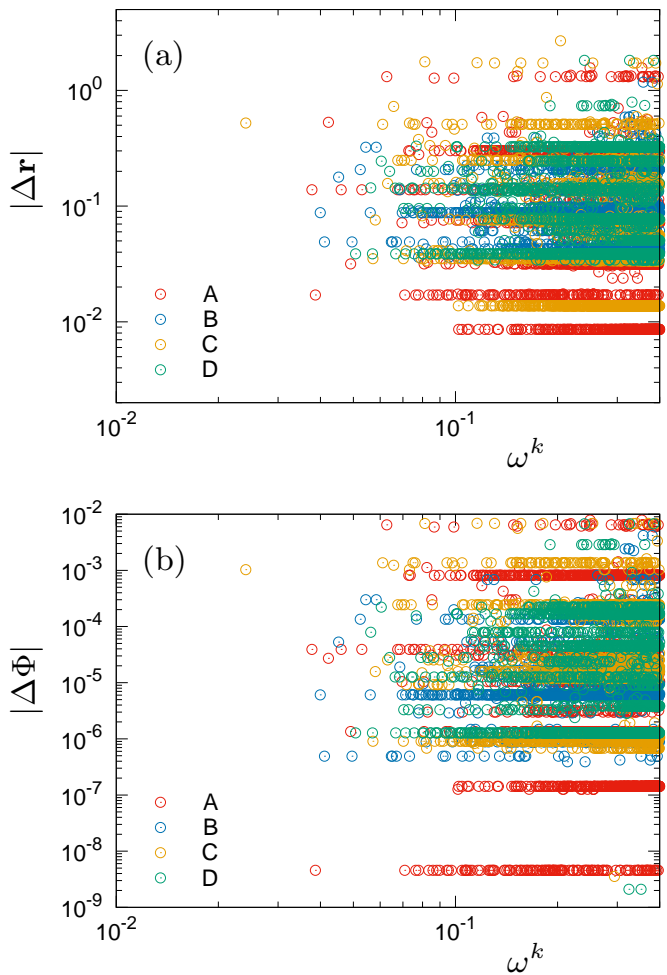}
\vspace*{0mm}
\caption{\label{fig12}
{Sample dependence of the transitions in the harmonic potential system.}
Plots of (a) $|\Delta \mathbf{r}|$ and (b) $|\Delta \Phi|$ as functions of the eigenfrequency of the excited vibrational mode, $\omega^k$.
The system size is $N = 16000$.
We plot data from different samples $A$ to $D$ using different color symbols.
}
\end{figure}

\subsection{Sample dependence}~\label{result:anha}
Finally, we study the sample dependence of the transitions.
Figure~\ref{fig12} plots (a) $|\Delta \mathbf{r}|$ and (b) $|\Delta \Phi|$ versus $\omega^k$ for four different samples of harmonic potential systems, $A$ to $D$, which are all composed of $N=16000$ particles.
We observe that the datasets of $|\Delta \mathbf{r}|$ and $|\Delta \Phi|$ fluctuate from sample to sample, and these fluctuations are rather random.
We therefore expect that in the thermodynamic limit of $N \to \infty$, $|\Delta \mathbf{r}|$ and $|\Delta \Phi|$ exhibit continuous distributions.
It is then important to study the functional forms of the probability distributions of $|\Delta \mathbf{r}|$ and $|\Delta \Phi|$.
In two-level system theory~\cite{Anderson_1972,Phillips_1972,Phillips_1987,Galperin_1989}, we assume a uniform distribution function for $|\Delta \Phi|$ in order to explain the linear temperature dependence of the specific heat.
This is beyond the present work but definitely an important future subject.

\section{Conclusion}
In this work, we have studied the anharmonic properties of the vibrational modes in two model amorphous solids, the harmonic potential system and LJ potential system.
Our results are summarized as follows.
(i)~The vibrational modes in amorphous solids exhibit strong anharmonicities that induce particle rearrangements and cause transitions to different states.
The onset temperature of anharmonicities is estimated as quite small values of $T_c \sim 10^{-9}$ for the harmonic potential system~($N=512000$) and $T_c \sim 10^{-3}$ for the LJ potential system~($N=128000$).
In the thermodynamic limit, $N \to \infty$, $T_c$ possibly goes to zero.
(ii)~Different vibrational modes do not always cause different transitions: The number of states after the transition is much fewer than that of vibrational modes.
(iii)~The induced particle rearrangements are always localized in space and include $1$ to $1000$ particles, with each particle's displacement being roughly $10^{-3}$ to $10^{-1}$ times the particle size.
This localized region is surrounded by some particle displacement field characterized by $|\Delta \mathbf{r}_i| \propto r^{-a}$~(with $a \approx 2.5$ to $3$).
(iv) The correlations between the anharmonic properties and nature of excited vibrational modes are rather weak.
Particularly, there are no apparent differences in the anharmonicities between the phonon modes and the QLV modes: both exhibit strong anharmonicities that cause the transitions.
(v)~In the thermodynamic limit of $N \to \infty$, we expect that the values of $|\Delta \mathbf{r}|$ and $|\Delta \Phi|$, which characterize the transitions, are continuously distributed.

The present results support the existence of TLS transitions in amorphous solids, which can correspond to localized rearrangements such as we observed.
Interestingly, experimental studies~\cite{Ruta_2012,Ruta_2014,Luo_2017} observed fast dynamics of atoms in the deeply glass state, which are distinct from aging dynamics.
Numerical simulations~\cite{Ozawa_2015,Ozawa_2018} also revealed the presence of localized excitations in randomly pinned glasses.
The present work detected a rather broad range of particle rearrangements composed of $1$ to $1000$ particles, with each particle's displacement being $10^{-3}$ to $10^{-1}$ times the particle size.
Further studies are needed to clarify which sizes of particle arrangements are related to the TLS transitions, which is determined by the energy barrier through the transition path connecting the two-level states~\cite{Weber_1985,Middleton_2001,Bonfanti_2017}.
Since the barrier of TLS transitions relevant at low $T$ is rather small, they could be related to only the small size of localized rearrangements~\cite{Jug_2016}.

In our recent work~\cite{Shimada2_2018}, we reported that the QLV modes exhibit unstable vibrations with negative vibrational energy in a localized region.
We might expect that the rearrangements occur along these unstable vibrations in the QLV modes.
However, although some rearrangements occur along the QLV modes, they generally do not.
We even found that the anharmonic properties are not relevant to the nature of excited vibrational modes.
We therefore argue that vibrational modes are just a trigger to induce TLS transitions, whereas the nature of the TLSs is determined by the complex shape of the energy landscape~\cite{Weber_1985,Heuer_1993,Heuer_1996,Reinisch_2004,Munro_1999,Middleton_2001,Damart_2018} that emerges due to the complex structural properties of amorphous systems~\cite{Hua_2018,Hua_2019,Tanaka_2019}.

In addition, we demonstrated that the extended phonon modes induce particle rearrangements and that even their extent of anharmonicity is similar to that of the QLV modes.
This implies that some defect-like vibrations are embedded in the phonon modes.
Indeed, Ref.~\cite{Wijtmans_2017} disentangled the localized defects from extended vibrational modes by implementing an artificial potential that acts as a high-pass filter.
Additionally, Refs.~\cite{Gartner_2016,Gartner2_2016} generated plastic modes (or nonlinear glassy modes, which are spatially localized) from extended modes by minimizing the energy-barrier function.
Importantly, these localized defects and plastic modes can be used to well predict the location of plastic instabilities when the system is mechanically deformed, i.e., they play a role as defects in the system.
We could expect that they also play a role in the transitions induced by thermal vibrations, although the transitions at low temperatures might be realized through quantum tunneling effects, which is another issue to be solved~\cite{Khomenko_2019}.

\section*{Acknowledgments}
We thank Walter~Kob for useful discussions and suggestions.
This work was supported by JSPS KAKENHI Grant Numbers 16H04034, 17H04853, 18H05225, 19J20036, 19H01812, and 19K14670.
This work was also supported by the Asahi Glass Foundation.
The theoretical calculations were partially performed using the Research Center for Computational Science, Okazaki, Japan.

\appendix

\begin{figure}[t]
\centering
\includegraphics[width=0.49\textwidth]{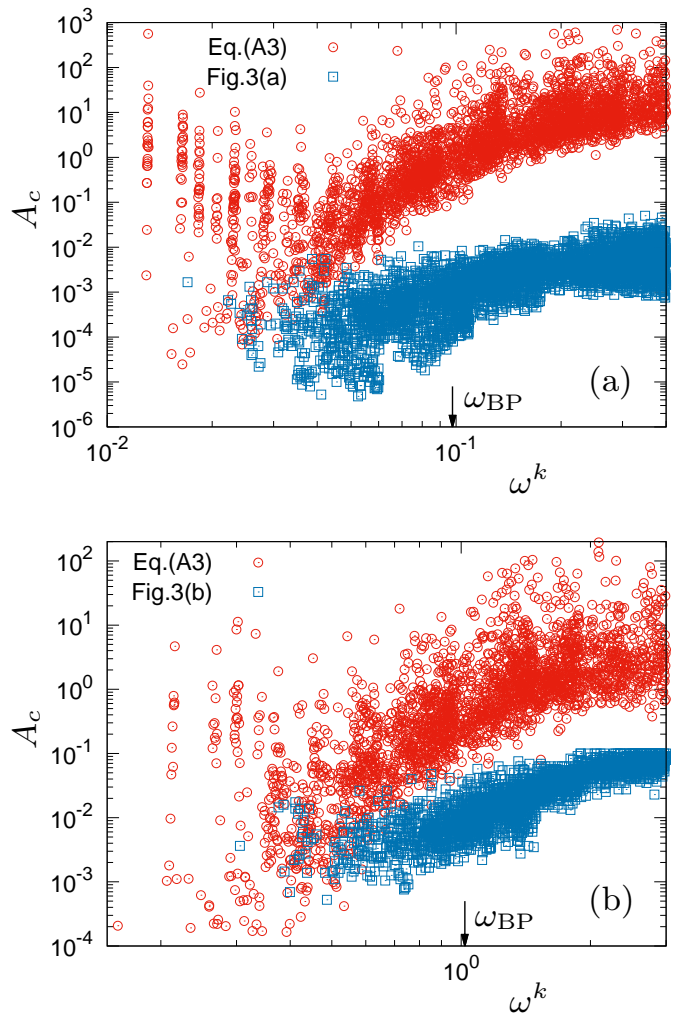}
\vspace*{0mm}
\caption{\label{fig13}
{Extent of anharmonicity of vibrational modes measured using Eq.~(\ref{acexpression}).}
Plot of the extent of anharmonicity, $A_c$, measured using Eq.~(\ref{acexpression}), as a function of the eigenfrequency of the excited vibrational mode, $\omega^k$.
(a) Harmonic potential system and (b) LJ potential system.
We plot data for different system sizes~(up to $N=4096000$ for the harmonic potential system and up to $N=1024000$ for the LJ potential system) together.
For comparison, we plot the values of $A_c$ presented in Fig.~\ref{fig3}.
Additionally, we indicate the boson peak frequency, $\omega_\text{BP}$, by the arrow.
}
\end{figure}

\begin{figure}[t]
\centering
\includegraphics[width=0.49\textwidth]{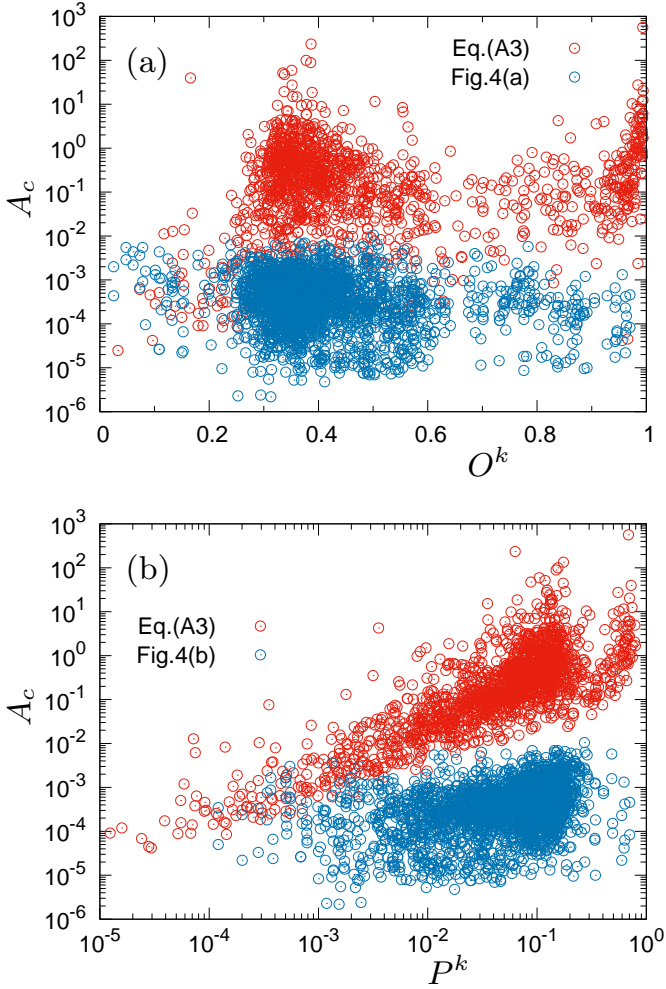}
\vspace*{0mm}
\caption{\label{fig14}
{Correlation between anharmonicities measured using Eq.~(\ref{acexpression}) and the nature of excited vibrational modes in the harmonic potential system.}
Plot of the extent of anharmonicity, $A_c$, measured using Eq.~(\ref{acexpression}), as a function of the phonon order parameter, $O^k$, in (a) and the participation ratio, $P^k$, in (b).
We plot data for vibrational modes below the boson peak frequency, $\omega_\text{BP}$.
For comparison, we plot the values of $A_c$ presented in Fig.~\ref{fig4}.
}
\end{figure}

\section{Expansion of the potential energy landscape around the inherent structure}~\label{expansion}
Here, we analyze the anharmonicities of the eigenmodes by expanding the potential energy, $\Phi(\{\mathbf{r}_i\})$, around the inherent structure $\{ \mathbf{r}_{0i} \}$ and along the direction of mode $k$~($\mathbf{e}^k_i$), as in Refs.~\cite{Gartner_2016,Gartner2_2016}.
We use $\{\mathbf{r}_i\}$ in Eq.~(\ref{eq:pull}) and expand $\Phi$ around the value of the inherent structure, $\Phi_0 \equiv \Phi(\{ \mathbf{r}_{0i} \})$, in terms of $A \sqrt{N}$ up to the third-order term:
\begin{equation}
\Delta \Phi = \Phi - \Phi_0 \approx \frac{1}{2} \kappa \left(A \sqrt{N} \right)^2 + \frac{1}{6} \tau \left(A \sqrt{N} \right)^3, 
\end{equation}
where
\begin{equation}
\begin{aligned}
\kappa &= \sum_{i,j=1}^N \frac{\partial^2 \Phi}{\partial \mathbf{r}_i \partial \mathbf{r}_j} : \mathbf{e}^k_i \mathbf{e}^k_j = {\omega^k}^2, \\
\tau   &= \sum_{i,j,k=1}^N \frac{\partial^3 \Phi}{\partial \mathbf{r}_i \partial \mathbf{r}_j \partial \mathbf{r}_k} \vdots \mathbf{e}^k_i \mathbf{e}^k_j \mathbf{e}^k_k.
\end{aligned}
\end{equation}
We then estimate the value of $A_c$ as the saddle point in the potential energy landscape of $\Phi(\{\mathbf{r}_i\})$:
\begin{equation}
A_c \sqrt{N} = -2 \frac{\kappa}{\tau} = -2 \frac{ {\omega^k}^2}{\tau}.  \label{acexpression}
\end{equation}
Note that $\tau$ should be negative.
We also emphasize that the value of $A_c$ in Eq.~(\ref{acexpression}) can be estimated using only the eigenvalues ${\omega^k}^2$~(second-order derivative of the potential) and the third-order derivative of the potential at $\{ \mathbf{r}_{i} \} = \{ \mathbf{r}_{0i} \}$.
We measure the values of $A_c$ in Eq.~(\ref{acexpression}) by employing larger system sizes, up to $N=4096000$ for the harmonic potential system and up to $N=1024000$ for the LJ potential system, and present them as a function of $\omega^k$ in Fig.~\ref{fig13} and as a function of $O^k$ and $P^k$ in Fig.~\ref{fig14}.
These results are discussed in Sec.~\ref{result:exanha} of the main text.

\bibliographystyle{apsrev4-1}
\bibliography{reference}

\end{document}